\documentclass[a4paper,11pt]{article}
\usepackage{pos}
\usepackage{bm}
\usepackage{slashed}

\def\be{\begin{equation}}
\def\ee{\end{equation}}
\def\ba{\begin{eqnarray}}
\def\ea{\end{eqnarray}}

\newcommand{\order}[1]{\mathcal{O}(#1)}

\title{Long-range interactions in double heavy tetraquarks $\bar Q \bar Q qq$}

\author*[a]{Muhammad Naeem Anwar}

\affiliation[a]{Department of Physics, Swansea University,
Singleton Park, Swansea, SA2 8PP, UK.}

\emailAdd{m.n.anwar@swansea.ac.uk}

\abstract{
At the large distances compared to the chiral symmetry breaking scale, a four-quark system $\bar Q \bar Q qq$ (with $Q$ as heavy and $q$ as light quarks) can be treated as two asymptotic mesons interacting via strong residual forces. The static heavy quark assumption enables using the Born-Oppenheimer approximation, where one can compute the potential between two heavy antiquarks in the presence of two light quarks of finite mass -- a prescription utilized in several lattice QCD studies. 
To analyse the long-range strong force in a $\bar Q \bar Q qq$ system, we study the interaction between two bottom mesons in the heavy quark limit using chiral effective field theory and dispersion theory with unphysical pion mass. We present methods to obtain two-pion-exchange potential between two static heavy mesons at non-physical pion mass and compare our preliminary results with the corresponding lattice QCD potentials.
}

\FullConference{The 41st International Symposium on Lattice Field Theory (LATTICE2024)\\
 28 July - 3 August 2024\\
Liverpool, UK\\}

\begin{document}
\maketitle

\section{Introduction}

The potential between two bottom mesons $B^{(*)} B^{(*)}$ in the heavy quark limit (HQL)\footnote{The $B$ and $B^*$ mesons are degenerated in the HQL, we use $B$ notation to represent this heavy-quark spin supermultiplet.} is a well-defined quantity due to the associated physical scale. In the HQL, the spin of the light degree of freedom becomes a good quantum number and most of the spin-dependent interactions are suppressed by the heavy quark mass, hence highly suppressed.

The Born-Oppenheimer (BO) approximation in the heavy quark limit provides a connection between the effective potential for the $\bar b \bar b q q$ system (four interacting quarks) and the $BB$ system (interacting mesons). The long-range QCD forces in $\bar b \bar b q q$ can be related to interacting $BB$ mesons through light-mesons exchanges. This brings the opportunity to investigate low-energy hadron-hadron interactions using chiral perturbation theory and dispersion theory to understand the interplay of mesons and quark degrees of freedom in $\bar b \bar b q q$ systems.

At short distances, the Coulomb-type strong interaction between two heavy quarks of the $BB$ system becomes dominant. However, at a large distance compared to the chiral symmetry breaking scale, the chiral potential is expected to be the dominant one.
The effective field theory that describes the interaction between two $B$ mesons is the same as between two nucleons. The rigorous analysis of the potential between two $B$ mesons will enlighten the possibility of shallow bound state(s) likewise deuteron in the nucleon sector. Hence, this exploratory study is helpful in understanding the nature of the strong force between heavy quarks and their multiquark bound states. 

The exploration of the existence of stable doubly-heavy tetraquarks has been an intriguing topic. The pioneer ChPT analysis was performed by Manohar and Wise~\cite{Manohar:1992nd}, and the earlier quark potential model exploration was reported in Ref.~\cite{Zouzou:1986qh}. The possibility of such stable states is also studied in chiral EFT~\cite{Wang:2018atz}, meson-exchange models~\cite{Li:2012ss,Liu:2019stu}, constituent quark model~\cite{Garcilazo:2020bgc}, and phenomenological approaches~\cite{Karliner:2017qjm,Eichten:2017ffp}.

There are several lattice QCD calculations of $\bar b \bar b q q$ system,  exploring the potential as a function of the distance between two static bottom quarks within the QCD and BO approximation~\cite{Detmold:2007wk,Bicudo:2015vta,Francis:2016hui,Bicudo:2017szl,Junnarkar:2018twb,Leskovec:2019ioa}. Their extracted binding energy of $\bar b \bar b q q$ system is around $\mathcal O(100)$ MeV, which is consistent with recent predictions of phenomenological models~\cite{Karliner:2017qjm,Eichten:2017ffp}.
In this study, we investigate the interactions at different ranges to understand the interplay between quark-level and hadron-level interactions (the so-called ``flip-flop" potential~\cite{Bicudo:2022cqi}).

The first principle studies such as Lattice QCD are instructive, however, performing Lattice QCD calculations for several different hadronic systems is computationally expensive. On the other hand, an EFT formalism calibrated to Lattice results for one hadronic state can be used to study other states related by QCD symmetries. One can extract the parameters of the EFT from the available Lattice data such as~\cite{Bicudo:2017szl} and exploit the symmetries of QCD (heavy quark spin and flavour symmetries) to make predictions for related states. Our objective here is to compute the chiral potentials between two heavy mesons by matching conditions to the lattice studies~\cite{Bicudo:2017szl} namely the static approximation and unphysical pion mass.

The main contributions to long and intermediate ranged $BB$ static potential would be:
\begin{itemize}
    \item One pion exchange, having range $ \mathcal O\left(1/m_\pi \right)$
    \item 2 pion exchange, driven by S-wave $\pi B$ interaction - range $ \mathcal O\left(1/2m_\pi \right)$
    \item 2 pion exchange, driven by P-wave $\pi B$ interaction - range $ \mathcal O\left(1/2m_\pi \right)$\,.
\end{itemize}

The subsystem interactions are very crucial here. The treatment of the 2 pion exchange potential at unphysical pion mass ($m_\pi=340$ MeV) is highly non-trivial due to the following reasons: (1) at this pion mass, the $f_0(500)/\sigma$ pole is very close to the 2$m_\pi$ threshold which makes $\pi \pi$ rescattering very essential to include~\cite{Hanhart:2008mx,Briceno:2016mjc}; 
(2) around $m_\pi=340$ MeV, the $\pi B$ S-wave interaction also develops a bound state, similar to the $D$-sector~\cite{Liu:2012zya}. There is additional $B^*$ left-hand-cut (LHC) to consider here. Therefore, our aim here is the proper inclusion of the subsystem interactions and investigate their interplay.

\section{Quantum numbers of $\bar b \bar b qq-BB$ system}

For a large $\bar b \bar b$ separations, the $\bar b \bar b qq$ system is approximated as $BB$ which is coupled in $S$-wave. The total parity of the system is positive which is the product of two negative parities of $B$ mesons. Assuming the flavour-symmetric $\bar b \bar b$ pair in colour $\bf 3$ (antisymmetric), Pauli principle ensures that the spin of the $\bar b \bar b$ pair, $j_b$, should be symmetric, hence $j_b=1$. For the light quark pair $qq$, there will be two possibilities:
\begin{enumerate}
\item
For flavour antisymmetric $qq=\dfrac{1}{\sqrt{2}}\Big(ud-du\Big)$, the isospin $I=0$ and spin of the light degree of freedom, $\vec{j_l}=\vec{L}+\vec{s}_{qq}$, must be antisymmetric, i.e. $j_l=0$.

\item
For flavour symmetric $qq=\Big(uu, dd, \frac{ud+du}{\sqrt{2}}\Big)$, the isospin $I=1$ and spin of the light degree of freedom must be symmetric, i.e. $j_l=1$.\end{enumerate}
All together, the possible combinations are: 

\begin{itemize}

\item
$qq=\dfrac{1}{\sqrt{2}}\Big(ud-du\Big)$, \big\{$I=0$, $j_l=0$, $j_b=1$ and $J=1$\big\}\\
$\Longrightarrow$ $I(J^P)=\Big\{0(1^+)\Big\}$

\item
$qq=\Big(uu, dd, \frac{ud+du}{\sqrt{2}}\Big)$, \big\{$I=1$, $j_l=1$, $j_b=1$ and $J=0,1,2$\big\}\\
$~~~~~~~~~~~~~~~~~~~~~~~~~~~~~~~~~~~~~~~~~~~~~~~~
~~~~~~~~~~~~\Longrightarrow$ $I(J^P)=\Big\{1(0^+),~1(1^+),~1(2^+)\Big\}$

\end{itemize}
We are interested in quantum numbers $J^P=1^+$, it is instructive to identify the physical asymptotic systems which can couple to this $J^P$, and these are $BB^*$ and $B^*B^*$ mesons pairs.

Bose symmetry further constrains the possible quantum numbers of two coupled mesons. In the heavy quark limit, it leads to the following selection rule;
\be
1=(-1)^{I+j_b+j_l+1}\,.
\ee
With fixed and conserved $j_b$, one can find the criteria when two asymptotic systems are allowed to host a bound state, i.e. $I=j_l$. This condition is supported by earlier lattice QCD explorations of the $BB$ potentials, where the attractive channels are found to be in $I=j_l$ configurations~\cite{Detmold:2007wk}, and the recent lattice Born-Oppenheimer potentials for $J^P=1^+$ also show the similar pattern~\cite{Bicudo:2017szl}.

\section{Heavy Meson $\chi$PT: One Pion Exchanges}

The leading order (LO) contribution to the potential between two $B$ mesons is the one pion exchange (OPE). The LO Lagrangian in HM$\chi$PT is
given by the kinetic energy of the heavy fields, the coupling of the heavy
fields to Goldstone bosons, and the mass splitting of the heavy
mesons~\cite{Wise:1992hn,Burdman:1992gh,Yan:1992gz}
\begin{eqnarray}\nonumber
\mathcal L_{\rm LO}&=&-i {\rm Tr}[\bar H_a v_\mu D^\mu_{ba} H_b]
+ g_\pi {\rm Tr}[\bar H_aH_b\gamma_\nu\gamma_5] u^\nu_{ba}\\
&&+\frac{\lambda}{m_Q} {\rm Tr}[\bar H_a\sigma_{\mu\nu}H_a\sigma^{\mu\nu}]
\end{eqnarray}
with the heavy fields defined as $H=\frac{1+\slashed{v}}{2}\left[\slashed{V}+iP\gamma_5\right],~\bar{\,H}=\gamma^0H^\dagger\gamma^0$,
where the trace is taken over the Gamma matrices and $a,b$ are the light flavour indices.
Here, the $P$ and $V_\mu$ stand for the pseudoscalar and vector heavy fields, respectively, and $v_\mu=(1,\vec{0})+\mathcal{O}(\vec p/M)$ is the heavy mesons velocity. In the heavy quark limit, the spin symmetry breaking term will vanish. Since heavy quark spin symmetry relates the couplings of the $V_\mu$ to those of the $P$, it is thus convenient to combine them into heavy meson superfield $H$.
The covariant derivative is
\begin{eqnarray}
D_\mu=\partial_\mu+\Gamma_\mu \, , \quad
\Gamma_\mu=\frac{1}{2}\left(u^\dagger\partial_\mu u+u\partial_\mu u^\dagger\right)
\end{eqnarray}
where $U=\exp\left(\sqrt{2}i\phi/F\right),~u^2=U$, with $F$ the pion decay constant in the chiral limit which later can be replaced by the physical pion decay constant $F_\pi$.
The Goldstone boson fields $\phi$ in matrix form are
\begin{eqnarray} 
&\phi= \left( \begin{array}{ccc}
\frac{1}{\sqrt{2}}\pi^0+\frac{1}{\sqrt{6}}\eta & \pi^+ & K^+ \\
\pi^- & -\frac{1}{\sqrt{2}}\pi^0+\frac{1}{\sqrt{6}}\eta &  K^0 \\
K^- & \bar{K}^0 & -\frac{2}{\sqrt{6}}\eta \end{array} \right).
\end{eqnarray}
The heavy meson axial coupling $g_\pi$ can be obtained from the well-measured $D^{*+} \to D^0 \pi^+$ decay.
In two component notation the heavy meson superfield is expressed as $H_a={\bm V}_a \cdot \boldsymbol{\sigma} + P_a$, with $\boldsymbol{\sigma}$ are the Pauli matrices.
The One Pion Exchange (OPE) potential can be obtained using the lowest order HM$\chi$PT Lagrangian for the axial coupling of the pion with the heavy mesons, viz.
\be
\mathcal{L}_{H\pi}=-\frac{g_\pi}{2} \langle H_{a}^\dagger H_b ~\boldsymbol{\sigma} \cdot \bm{u}_{ab} \rangle\, ,
\ee
where $\bm{u}=-\sqrt{2} \boldsymbol{\nabla} \phi / F_\pi$.
The OPE potential between two $B$ mesons in the momentum space is
\be
V_{\pi}(q)=(I_1 \cdot I_2) ~ \frac{g_\pi^2}{F_{\pi}^2}~ \frac{(q \cdot \epsilon_2)(q \cdot \epsilon_4^*)}{q^2-m_{\pi}^2} \, .
\ee
In the position space, the OPE potential in $S$-wave is given by
\be
V_{\pi}(r)=(I_1 \cdot I_2) ~ \frac{g_\pi^2}{3F_{\pi}^2} ~ \bigg(m_\pi^2 ~ \frac{e^{-m_\pi r}}{4 \pi r} - \delta(r) \bigg) \, ,
\ee
with $I_1 \cdot I_2=-\dfrac{3}{4}$ for $I=0$ and $I_1 \cdot I_2 =~\dfrac{1}{4}$ for $I=1$.

\begin{figure}[h!]
    \centering
    \includegraphics[width=0.6\textwidth]{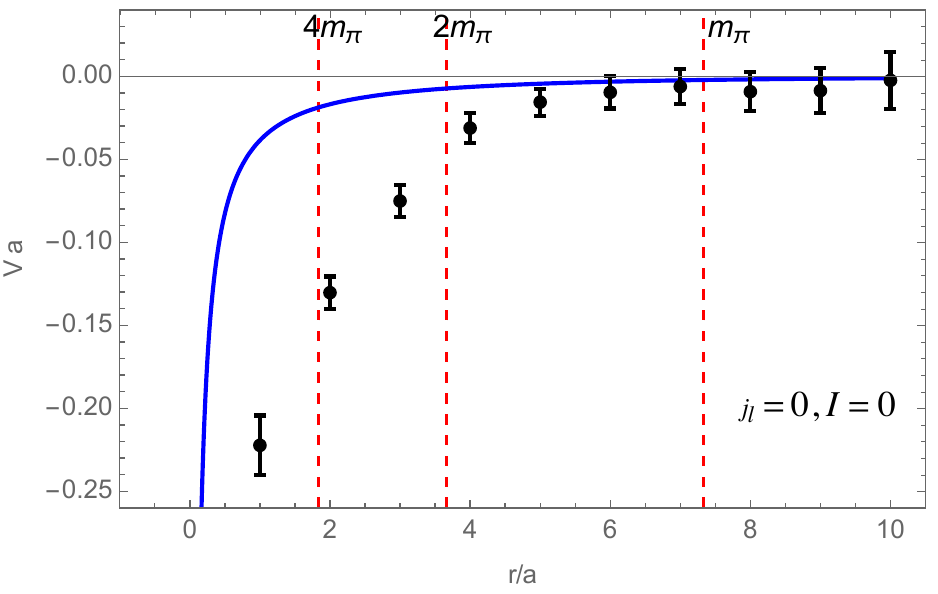}
    \caption{OPE potential in lattice units with lattice data from~\cite{Bicudo:2017szl} with $m_{\pi}=340$ MeV.}
    \label{OPE}
\end{figure}

The OPE potential (in lattice units) for the $s_l=0, I=0$ configuration is shown in the Fig.~\ref{OPE} along with lattice data~\cite{Bicudo:2017szl}. This channel is found to be the most attractive in the lattice calculations. Parameters are as following: $m_\pi=340$ MeV, $f_\pi(m_\pi=340)=114$ MeV, and $g_\pi=0.5$.
The long tail of the OPE potential is clearly visible in the lattice potential.

\section{Two pion exchange potential}

\begin{figure}[h]
    \centering
    \includegraphics[width=0.8\textwidth]{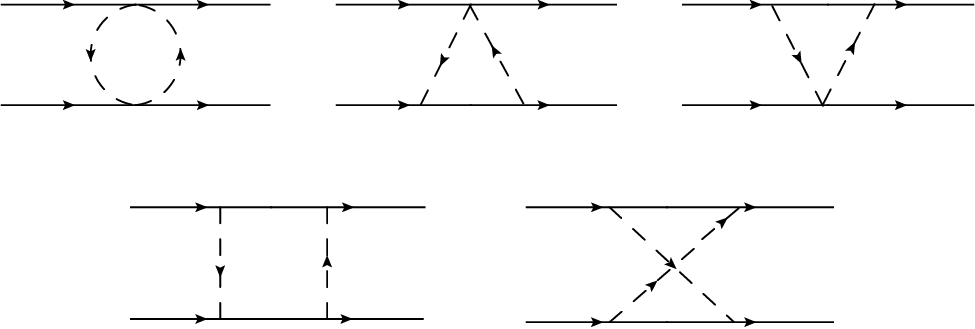}
    \caption{Two pion exchanges between two bottom mesons. The sold lines indicate the $B$ mesons (in the heavy quark limit) and dashed lines represent pions.}
    \label{2PEdiag}
\end{figure}

The possible diagrams for the two pion exchange (2PE) potential are sketched in Fig.~\ref{2PEdiag}.

2PE potential can be calculated by appropriately multiplying together the relevant $B\pi$ scattering diagrams, similar to as nucleon-nucleon potentials~\cite{Donoghue:2006rg}. For the correlated $\pi \pi$ subsystem, unitarity requires the inclusion of the $\pi \pi$ rescattering~\cite{Hanhart:2008mx} and we use the dispersion approach to include it.

\subsection{Dispersion Approach}

The 2PE potential can be obtained using the dispersion relation,
\be
\mathcal M(s,t)=\frac1\pi \int\limits_{4m_\pi^2}^{\infty} dz  \frac{{\rm Im} {\mathcal M}(z,t)}{z-s-i\epsilon}
\ee
To evaluate the discontinuity in the above amplitude, consider $\Gamma_{\rm out}=1+T_{\pi\pi}G$ be the $\pi\pi$ vertex function, where $T_{\pi\pi}$ denotes the $\pi\pi$ scattering matrix and $G$ is the 2$\pi$ propagator. In general, one can write
\be
{\rm disc}~\Gamma_{\rm out}= 2 i \sigma T^* \Gamma_{\rm out}
\ee
where $\sigma=\dfrac{\sqrt{1-4m_\pi^2/s}}{16\pi}$ with $s$ is the $\pi \pi$ invariant mass. With this we can write for the $B\bar B \to \pi\pi \to B\bar B$ transition amplitude
\be
\mathcal M_{\pi\pi}= G ~\Gamma_{\rm out} a^2
\ee
where $a$ is $B\bar B \to \pi\pi$ transition strength. Following~\cite{Hanhart:2012wi}, the discontinuity of the $\mathcal M_{\pi\pi}$
can be written as
\be
{\rm Im} \mathcal M_{\pi\pi}= 2 i \sigma a^2 |\Gamma_{\rm out}|^2 \,.
\ee
\begin{figure}[t!]
    \centering
    \includegraphics[width=0.35\textwidth]{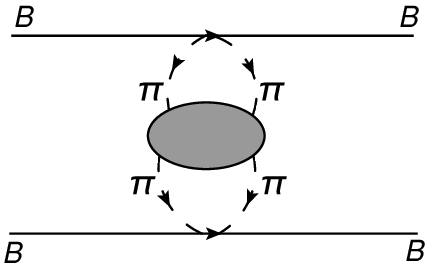}
    \caption{Correlated two pion exchange between two bottom mesons.}
    \label{2PE}
\end{figure}
First, let us focus on the $S$-wave $\pi\pi$ scattering in isospin $I=0$ channel, namely the one saturated by $\sigma$ meson exchange~\cite{Fujii:1999xn}. We may identify
\be
\Gamma_{\rm out}=\Omega^{L=0}(s)={\rm exp} \Big[\frac{s}{\pi} \int\limits_{4m_\pi^2}^{\infty} dz \frac{\delta^0(z)}{z(z-s-i\epsilon)}\Big]
\ee
To fix $a$, we identify the $B\bar B \to \pi\pi$ transition amplitude with the corresponding $\pi B \to \pi B$ scattering length. This gives
\be
(a^0)^2=\sum_{\alpha,\beta} a^{(+)2} \delta_{\alpha\beta} \delta_{\alpha\beta}=3 T^{(+)2}\, .
\ee
Here we use the following isospin basis relation.
The different isospin basis can be related to the operator representation via
\be
T_{\alpha \beta}=T^{(+)} \delta_{\alpha \beta} -T^{(-)}\frac12 \left[\tau_\alpha, \tau_\beta \right]
\label{Ibasis}
\ee
with $T^{(+)}=\frac13 \left( T^{\frac12}+2T^{\frac32}\right)$, and $T^{(-)}= \frac13 \left( -T^{\frac12}+T^{\frac32}\right)$.
All together, the discontinuity of the $\mathcal M_{\pi\pi}$ for the scalar-isoscalar channel is
\be
{\rm Im} \mathcal M_{\pi\pi}^{I=0}= -\frac32 i \pi \left(a^{(+)}\right)^2 \sqrt{1-\frac{4m_\pi^2}{s}} \left|\Omega^{0}(s)\right|^2 \Theta(s-4m_\pi^2)
\label{disc}
\ee
We need to evaluate the $\pi B$ scattering lengths $a^+$ to proceed.

\subsection{Scattering Length Approximation}

The $\pi B $ scattering lengths can be evaluated by using next-to leading order (NLO)  $\pi H$ scattering potentials~\cite{Liu:2012zya}. The $\pi B$ scattering amplitudes are given by
\be
\label{eq:v}
V(s,t,u) = \frac1{F_\pi^2} \bigg[\frac{C_{\rm LO}}{4}(s-u) - 4 C_0 h_0 +
2 C_1 h_1 - 2C_{24} H_{24}(s,t,u) + 2C_{35} H_{35}(s,t,u) \bigg],
\ee
where $h_i$ are low-energy-constants (LEC) and $C_i$ are the coefficients that can be found in~\cite{Liu:2012zya}. 
At $\pi H$ threshold, we get
\be
V^{H\pi}_{\rm{thr}} = \frac{m_\pi^2}{f_\pi^2} \bigg[C_{\rm LO}\frac{M_H}{m_\pi} - 4 h_0 -
2 h_1 - 4(h_2+h_4 M_H^2) +2(h_3+ 2h_5 M_H^2) \bigg]\,.
\ee
It is clear to see that all these terms have the same $m_Q$ scaling: $h_{0,1,24,35} \sim \order{m_Q}=\order{M_H}$. Thus we may write the LECs as
\be
h_{0,1,24,35}^Q=h_{0,1,24,35}^c ~\frac{M_H}{M_D}\,.
\ee
This now relates the LEC at arbitrary heavy quark $Q$ sector to those of determine in the charm sector. The NLO scattering amplitude at threshold becomes
\be
V^{H\pi,I}_{\rm{thr}} = \frac{m_\pi^2 M_H}{f_\pi^2} \bigg[\frac{C_{\rm LO}^I}{m_\pi} - \frac2{M_D}\Big(2h_0^c + h_1^c + 2h_{24}^c -h_{35}^c \Big) \bigg]\,.
\ee
The two particle loop function using a hard cut-off in 3-momentum at threshold is evaluated as  
\begin{eqnarray}\nonumber
G^\Lambda_{\rm{thr}}&=&
\frac{1}{16\pi^2
  (M_H+m_\pi)} \left[M_H \log\left(\frac{M_H^2}{\big(\sqrt{M_H^2+\Lambda^2}+\Lambda^2 \big)^2}\right)+m_\pi \log\left(\frac{m_\pi^2}{\big(\sqrt{m_\pi^2+\Lambda^2}+\Lambda^2 \big)^2}\right)\right]\,.
\label{nonrelloop}
\end{eqnarray}
It is clear to see that the loop function $G^\Lambda_{\rm{threshold}} \sim \order{1/M_H}$.
The hard cutoff can set by matching with the dimensional regularization subtraction constant as~\cite{Guo:2005wp}. In the charm sector, the best fit leads to $\Lambda \sim 700$ MeV.
Heavy-meson$-$pion scattering length can be obtained by using the unitarized scattering amplitude~\cite{Guo:2009ct}
\be
a^I_0=-\frac{M_H T_{\rm thr,NR}^I}{4\pi(M_H+m\pi)} \,
\quad \text{with} \quad
T_{\rm thr}^I= \frac{V^{H\pi,I}_{\rm{thr}}}{\big[1-V^{H\pi, I}_{\rm{thr}}G^\Lambda_{\rm{thr}}\big]} \,.
\label{sla0}
\ee
Since we are dealing with infinitely heavy quarks, above amplitude are nonrelativistic with proper normalization as $T_{\rm thr,Rel}^I=2M_H T_{\rm thr,NR}^I$.
The LECs $h_i$'s were extracted from the fit to the pion mass dependence of a set of charmed-meson--light-meson scattering lengths performed in Ref.~\cite{Liu:2012zya}.

\subsection{Two Pion Exchange Potential via Spectral Function}

The long range two pion exchange potential in the $t$-channel is saturated by $\sigma$ meson exchange which can be constructed dispersively assuming that the $B \bar B \to \pi \pi$ transition amplitude has no left hand cuts (LHCs). Following Donoghue~\cite{Donoghue:2006rg}, the momentum space potential of exchanging $S$-wave $\pi \pi$ is
\be
V_\sigma (q^2)= \frac2\pi \int\limits_{2m_\pi}^{\infty} d\mu~\mu \frac{{\rm Im} {\mathcal M}(s,\mu^2)}{\mu^2+q^2}
\ee
and, in the position space
\be
V_\sigma (r)= \frac1{2 \pi^2 r} \int\limits_{2m_\pi}^{\infty} d\mu~\mu e^{-\mu r}{\rm Im} {\mathcal M}(s,\mu^2) \,.
\ee
It is thus clear that to get the potential we need compute the ${\rm Im} {\mathcal M}(s,t)$.
We have settled this in the previous subsections. The discontinuity of the $\mathcal M_{\pi\pi}$
for the scalar-isoscalar channel is given in Eq.\,(\ref{disc}).
Finally, the potential for the $\sigma$ exchange using the Omn\`es representation is
\begin{equation}
  V_\sigma(r) = - \frac{3}{4\pi r}\left( a^{(+)} \right)^2 \int_{2m_\pi}^\infty d\mu 
    \sqrt{\mu^2-4m_\pi^2} e^{-\mu r} \left|\Omega^0(\mu^2)\right|^2,
\end{equation}
with $a^{(+)} = -0.54^{+0.19}_{-0.59}$~fm is computed using Eq.\,(\ref{sla0}) and in the static limit.
The error is propagated from the uncertainties of the LECs $h_{24}^c$ and $h_{35}^c$, and the nonlinear dependence of the scattering lengths on the LECs is taken into account in the error propagation by computing the scattering lengths using 150 sets of LECs at the edge of the $1\sigma$ region of the fit done in Ref.~\cite{Liu:2012zya}.

\begin{figure}[h]
    \centering
    \includegraphics[width=0.6\textwidth]{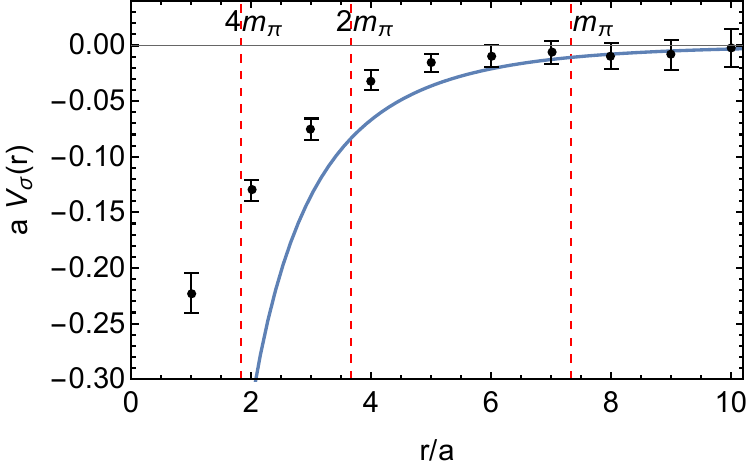}
    \caption{2PE potential (preliminary) for the $s_l=0, I=0$ channel in the scattering length approximation.}
    \label{2PEpot}
\end{figure}
\noindent
The 2PE potential (preliminary) for $s_l=0, I=0$ channel is shown in lattice units in Fig.~\ref{2PEpot} along with lattice data~\cite{Bicudo:2017szl}. Parameters are: $m_\pi=340$\,MeV, $f_\pi(m_\pi$=$340)=114$ MeV~\cite{Bicudo:2017szl}, and $g_\pi=0.5$. One can identify the onsets of the 2PE potential in the lattice data around the $2m_\pi$ distances.

\section{Conclusion}

This work elucidates methods for obtaining the chiral potentials between two heavy mesons in the static limit at heavy pion mass. We compare our preliminary results (without left-hand cuts) with the lattice data. The obtained chiral potentials clearly describe the long-range lattice potential well. The intermediate range needs a more rigorous treatment.

In the future, we plan to include all possible LHCs and utilize the Khuri-Treiman formalism to obtain intermediate-range interactions in different spin and isospin channels such as correlated $\rho$ and $\omega$  mesons exchanges~\cite{Anwar:2025}. In addition to improving formalism, we plan to extend our analysis to the updated lattice data presented at the Lattice 2024 conference~\cite{Bicudo:2024vxq}.

\section*{Acknowledgement}
We are grateful to Feng-Kun Guo, Christoph Hanhart, Bastian Kubis, Thomas C. Luu, and Ulf-G. Mei\ss ner for stimulating discussions and invaluable mentoring, and to Pedro Bicudo and Marc Wagner for insightful discussions and providing their lattice data of Ref.~\cite{Bicudo:2017szl}. This work is supported by The Royal Society through Newton International Fellowship.

\bibliographystyle{JHEP}
\bibliography{skeleton}

\providecommand{\href}[2]{#2}\begingroup\raggedright\begin{thebibliography}{10}

\bibitem{Manohar:1992nd}
A.V.~Manohar and M.B.~Wise, \emph{{Exotic Q Q anti-q anti-q states in QCD}}, \href{https://doi.org/10.1016/0550-3213(93)90614-U}{\emph{Nucl. Phys. B} {\bfseries 399} (1993) 17} [\href{https://arxiv.org/abs/hep-ph/9212236}{{\ttfamily hep-ph/9212236}}].

\bibitem{Zouzou:1986qh}
S.~Zouzou, B.~Silvestre-Brac, C.~Gignoux and J.M.~Richard, \emph{{Four-quark bound states}}, \href{https://doi.org/10.1007/BF01557611}{\emph{Z. Phys. C} {\bfseries 30} (1986) 457}.

\bibitem{Wang:2018atz}
B.~Wang, Z.-W.~Liu and X.~Liu, \emph{{$\bar{B}^{(\ast)} \bar{B}^{(\ast)}$ interactions in chiral effective field theory}}, \href{https://doi.org/10.1103/PhysRevD.99.036007}{\emph{Phys. Rev. D} {\bfseries 99} (2019) 036007} [\href{https://arxiv.org/abs/1812.04457}{{\ttfamily 1812.04457}}].

\bibitem{Li:2012ss}
N.~Li, Z.-F.~Sun, X.~Liu and S.-L.~Zhu, \emph{{Coupled-channel analysis of the possible $D^{(*)}D^{(*)}, \overline{B}^{(*)}\overline{B}^{(*)}$ and $D^{(*)}\overline{B}^{(*)}$ molecular states}}, \href{https://doi.org/10.1103/PhysRevD.88.114008}{\emph{Phys. Rev. D} {\bfseries 88} (2013) 114008} [\href{https://arxiv.org/abs/1211.5007}{{\ttfamily 1211.5007}}].

\bibitem{Liu:2019stu}
M.-Z.~Liu, T.-W.~Wu, M.~Pavon~Valderrama, J.-J.~Xie and L.-S.~Geng, \emph{{Heavy-quark spin and flavor symmetry partners of the X(3872) revisited: What can we learn from the one boson exchange model?}}, \href{https://doi.org/10.1103/PhysRevD.99.094018}{\emph{Phys. Rev. D} {\bfseries 99} (2019) 094018} [\href{https://arxiv.org/abs/1902.03044}{{\ttfamily 1902.03044}}].

\bibitem{Garcilazo:2020bgc}
H.~Garcilazo and A.~Valcarce, \emph{{Hidden and Open Heavy-Flavor Hadronic States}}, \href{https://doi.org/10.1007/s00601-020-01557-1}{\emph{Few Body Syst.} {\bfseries 61} (2020) 24} [\href{https://arxiv.org/abs/2007.06046}{{\ttfamily 2007.06046}}].

\bibitem{Karliner:2017qjm}
M.~Karliner and J.L.~Rosner, \emph{{Discovery of doubly-charmed $\Xi_{cc}$ baryon implies a stable ($b b \bar{u} \bar{d}$) tetraquark}}, \href{https://doi.org/10.1103/PhysRevLett.119.202001}{\emph{Phys. Rev. Lett.} {\bfseries 119} (2017) 202001} [\href{https://arxiv.org/abs/1707.07666}{{\ttfamily 1707.07666}}].

\bibitem{Eichten:2017ffp}
E.J.~Eichten and C.~Quigg, \emph{{Heavy-quark symmetry implies stable heavy tetraquark mesons $Q_iQ_j \bar q_k \bar q_l$}}, \href{https://doi.org/10.1103/PhysRevLett.119.202002}{\emph{Phys. Rev. Lett.} {\bfseries 119} (2017) 202002} [\href{https://arxiv.org/abs/1707.09575}{{\ttfamily 1707.09575}}].

\bibitem{Detmold:2007wk}
W.~Detmold, K.~Orginos and M.J.~Savage, \emph{{BB Potentials in Quenched Lattice QCD}}, \href{https://doi.org/10.1103/PhysRevD.76.114503}{\emph{Phys. Rev. D} {\bfseries 76} (2007) 114503} [\href{https://arxiv.org/abs/hep-lat/0703009}{{\ttfamily hep-lat/0703009}}].

\bibitem{Bicudo:2015vta}
P.~Bicudo, K.~Cichy, A.~Peters, B.~Wagenbach and M.~Wagner, \emph{{Evidence for the existence of $u d \bar{b} \bar{b}$ and the non-existence of $s s \bar{b} \bar{b}$ and $c c \bar{b} \bar{b}$ tetraquarks from lattice QCD}}, \href{https://doi.org/10.1103/PhysRevD.92.014507}{\emph{Phys. Rev. D} {\bfseries 92} (2015) 014507} [\href{https://arxiv.org/abs/1505.00613}{{\ttfamily 1505.00613}}].

\bibitem{Francis:2016hui}
A.~Francis, R.J.~Hudspith, R.~Lewis and K.~Maltman, \emph{{Lattice Prediction for Deeply Bound Doubly Heavy Tetraquarks}}, \href{https://doi.org/10.1103/PhysRevLett.118.142001}{\emph{Phys. Rev. Lett.} {\bfseries 118} (2017) 142001} [\href{https://arxiv.org/abs/1607.05214}{{\ttfamily 1607.05214}}].

\bibitem{Bicudo:2017szl}
P.~Bicudo, M.~Cardoso, A.~Peters, M.~Pflaumer and M.~Wagner, \emph{{$u d \bar{b} \bar{b}$ tetraquark resonances with lattice QCD potentials and the Born-Oppenheimer approximation}}, \href{https://doi.org/10.1103/PhysRevD.96.054510}{\emph{Phys. Rev. D} {\bfseries 96} (2017) 054510} [\href{https://arxiv.org/abs/1704.02383}{{\ttfamily 1704.02383}}].

\bibitem{Junnarkar:2018twb}
P.~Junnarkar, N.~Mathur and M.~Padmanath, \emph{{Study of doubly heavy tetraquarks in Lattice QCD}}, \href{https://doi.org/10.1103/PhysRevD.99.034507}{\emph{Phys. Rev. D} {\bfseries 99} (2019) 034507} [\href{https://arxiv.org/abs/1810.12285}{{\ttfamily 1810.12285}}].

\bibitem{Leskovec:2019ioa}
L.~Leskovec, S.~Meinel, M.~Pflaumer and M.~Wagner, \emph{{Lattice QCD investigation of a doubly-bottom $\bar{b} \bar{b} u d$ tetraquark with quantum numbers $I(J^P) = 0(1^+)$}}, \href{https://doi.org/10.1103/PhysRevD.100.014503}{\emph{Phys. Rev. D} {\bfseries 100} (2019) 014503} [\href{https://arxiv.org/abs/1904.04197}{{\ttfamily 1904.04197}}].

\bibitem{Bicudo:2022cqi}
P.~Bicudo, \emph{{Tetraquarks and pentaquarks in lattice QCD with light and heavy quarks}}, \href{https://doi.org/10.1016/j.physrep.2023.10.001}{\emph{Phys. Rept.} {\bfseries 1039} (2023) 1} [\href{https://arxiv.org/abs/2212.07793}{{\ttfamily 2212.07793}}].

\bibitem{Hanhart:2008mx}
C.~Hanhart, J.R.~Pelaez and G.~Rios, \emph{{Quark mass dependence of the rho and sigma from dispersion relations and Chiral Perturbation Theory}}, \href{https://doi.org/10.1103/PhysRevLett.100.152001}{\emph{Phys. Rev. Lett.} {\bfseries 100} (2008) 152001} [\href{https://arxiv.org/abs/0801.2871}{{\ttfamily 0801.2871}}].

\bibitem{Briceno:2016mjc}
R.A.~Briceno, J.J.~Dudek, R.G.~Edwards and D.J.~Wilson, \emph{{Isoscalar $\pi\pi$ scattering and the $\sigma$ meson resonance from QCD}}, \href{https://doi.org/10.1103/PhysRevLett.118.022002}{\emph{Phys. Rev. Lett.} {\bfseries 118} (2017) 022002} [\href{https://arxiv.org/abs/1607.05900}{{\ttfamily 1607.05900}}].

\bibitem{Liu:2012zya}
L.~Liu, K.~Orginos, F.-K.~Guo, C.~Hanhart and U.-G.~Meissner, \emph{{Interactions of charmed mesons with light pseudoscalar mesons from lattice QCD and implications on the nature of the $D_{s0}^*(2317)$}}, \href{https://doi.org/10.1103/PhysRevD.87.014508}{\emph{Phys. Rev. D} {\bfseries 87} (2013) 014508} [\href{https://arxiv.org/abs/1208.4535}{{\ttfamily 1208.4535}}].

\bibitem{Wise:1992hn}
M.B.~Wise, \emph{{Chiral perturbation theory for hadrons containing a heavy quark}}, \href{https://doi.org/10.1103/PhysRevD.45.R2188}{\emph{Phys. Rev. D} {\bfseries 45} (1992) R2188}.

\bibitem{Burdman:1992gh}
G.~Burdman and J.F.~Donoghue, \emph{{Union of chiral and heavy quark symmetries}}, \href{https://doi.org/10.1016/0370-2693(92)90068-F}{\emph{Phys. Lett. B} {\bfseries 280} (1992) 287}.

\bibitem{Yan:1992gz}
T.-M.~Yan, H.-Y.~Cheng, C.-Y.~Cheung, G.-L.~Lin, Y.C.~Lin and H.-L.~Yu, \emph{{Heavy quark symmetry and chiral dynamics}}, \href{https://doi.org/10.1103/PhysRevD.46.1148}{\emph{Phys. Rev. D} {\bfseries 46} (1992) 1148}.

\bibitem{Donoghue:2006rg}
J.F.~Donoghue, \emph{{Sigma exchange in the nuclear force and effective field theory}}, \href{https://doi.org/10.1016/j.physletb.2006.10.033}{\emph{Phys. Lett. B} {\bfseries 643} (2006) 165} [\href{https://arxiv.org/abs/nucl-th/0602074}{{\ttfamily nucl-th/0602074}}].

\bibitem{Hanhart:2012wi}
C.~Hanhart, \emph{{A New Parameterization for the Pion Vector Form Factor}}, \href{https://doi.org/10.1016/j.physletb.2012.07.038}{\emph{Phys. Lett. B} {\bfseries 715} (2012) 170} [\href{https://arxiv.org/abs/1203.6839}{{\ttfamily 1203.6839}}].

\bibitem{Fujii:1999xn}
H.~Fujii and D.~Kharzeev, \emph{{Long range forces of QCD}}, \href{https://doi.org/10.1103/PhysRevD.60.114039}{\emph{Phys. Rev. D} {\bfseries 60} (1999) 114039} [\href{https://arxiv.org/abs/hep-ph/9903495}{{\ttfamily hep-ph/9903495}}].

\bibitem{Guo:2005wp}
F.-K.~Guo, R.-G.~Ping, P.-N.~Shen, H.-C.~Chiang and B.-S.~Zou, \emph{{S wave K pi scattering and effects of kappa in J/psi ---\ensuremath{>} anti-K*0 (892) K+ pi-}}, \href{https://doi.org/10.1016/j.nuclphysa.2006.04.008}{\emph{Nucl. Phys. A} {\bfseries 773} (2006) 78} [\href{https://arxiv.org/abs/hep-ph/0509050}{{\ttfamily hep-ph/0509050}}].

\bibitem{Guo:2009ct}
F.-K.~Guo, C.~Hanhart and U.-G.~Meissner, \emph{{Interactions between heavy mesons and Goldstone bosons from chiral dynamics}}, \href{https://doi.org/10.1140/epja/i2009-10762-1}{\emph{Eur. Phys. J. A} {\bfseries 40} (2009) 171} [\href{https://arxiv.org/abs/0901.1597}{{\ttfamily 0901.1597}}].

\bibitem{Anwar:2025}
M.N.~Anwar and at. al., \emph{Deciphering chiral physics in lattice born-openheimer potentials}, {\emph{in preparation} (2025) }.

\bibitem{Bicudo:2024vxq}
P.~Bicudo, M.~Krstic~Marinkovic, L.~M\"uller and M.~Wagner, \emph{{Antistatic-antistatic $\bar Q \bar Q qq$ potentials for $u$, $d$ and $s$ light quarks from lattice QCD}}, \href{https://doi.org/10.22323/1.466.0124}{\emph{PoS} {\bfseries LATTICE2024} (2025) 124} [\href{https://arxiv.org/abs/2409.10786}{{\ttfamily 2409.10786}}].

\end{thebibliography}\endgroup

\end{document}